# Building Custom Term Suggestion Web Services with OAI-Harvested Open Data

Philipp Schaer, Thomas Lüke, Wilko van Hoek

GESIS – Leibniz Institute for the Social Sciences, Unter Sachsenhausen 6-8, 50667 Köln

{philipp.schaer, thomas.lueke, wilko.vanhoek}@gesis.org

## Introduction

With the rise of large web search engines the problem of "finding the right words" to express an information need lost its former important role in information retrieval. Users will get results for whatever query term they use – thanks to the remarkably vast document indexes. Despite the disputable quality of these retrieval, empty result lists seems like a phenomenon from a long forgotten era. With the growing success of freely available scientific open access material on the web the long-known "vocabulary problem" (Furnas, Landauer, Gomez, & Dumais, 1987) gets relevant again.

The problem that the same information need can be expressed in a variety of ways is especially true for scientific literature. Each scientific discipline has its own domain-specific language and vocabulary. This language is coded into documentary tools like thesauri or classifications that are used to document and describe scientific documents. When we think of information retrieval as "fundamentally a linguistic process" (Blair, 2003) users have to be aware of the most relevant search terms – which are the controlled thesauri terms the documents are described with. This can be achieved with so-called search-term-recommenders (STR) that map free search terms of a lay user to controlled vocabulary terms which can then be used as a term suggestion or to do an automatic query expansion (Hienert, Schaer, Schaible, & Mayr, 2011).

State-of-the-art repository software systems like DSpace or EPrints already offer some kind of term suggestion features in search or input forms but these implementations only work as simple auto completion mechanisms that don't incorporate any kind of semantic mapping. Such software systems would gain a lot in terms of usability and data consistency if tools like the proposed domain-specific STRs would be freely available. We aim to implement a rich toolbox of web services (like the mentioned domain-specific STRs) to support users and providers of online Digital Library (DL) or repository systems.

In this paper we will present an overall approach to implement such a STR on the basis of freely available open access and open metadata sets that are accessible via the well-known Open Archives Initiative Protocol for Metadata Harvesting (OAI-PMH)[1] interface. We will show how repositories can be processed to create a web service and what API is used to implement such a service into custom DL systems.

---

[1] http://www.openarchives.org/OAI/openarchivesprotocol.html

# Use Cases and Application Examples

A domain-specific STR offers various forms of usage. In the following section we will describe use-cases in which providers of DL systems may improve their query system through our domain-specific STR web service. The following scenarios will be presented:

- Search-Term-Recommendation for manual query expansion,
- automatic query expansion,
- tag cloud generation,
- bibliometric analysis.

## Search-Term-Recommendation for Manual Query Expansion

While users may know what they are searching for, they might not know how to phrase this information need. A user looking for articles related to "unemployment of young people" could benefit from expanding the query with domain-specific words. Let us assume that most of the relevant documents for this user's query are annotated with terms from a controlled vocabulary, e.g. a thesaurus. While "young people" might not be part of this thesaurus, terms like "labour market policy" or "training position" might be. Expanding the query with such terms will give the user a different result set. Since those terms are semantically related to the original search terms it is also likely that the additional documents in the result set are relevant to the user (Mutschke, Mayr, Schaer, & Sure, 2011).

Providers of DL systems can integrate the recommendations into their search form. While typing a query several related terms are presented to the user. These related terms could be added to the query. Figure 1 shows an example of this for the information platform for social sciences Sowiport[2]. Since the domain specific data in Sowiport is mostly German the recommendations in the example are German as well.

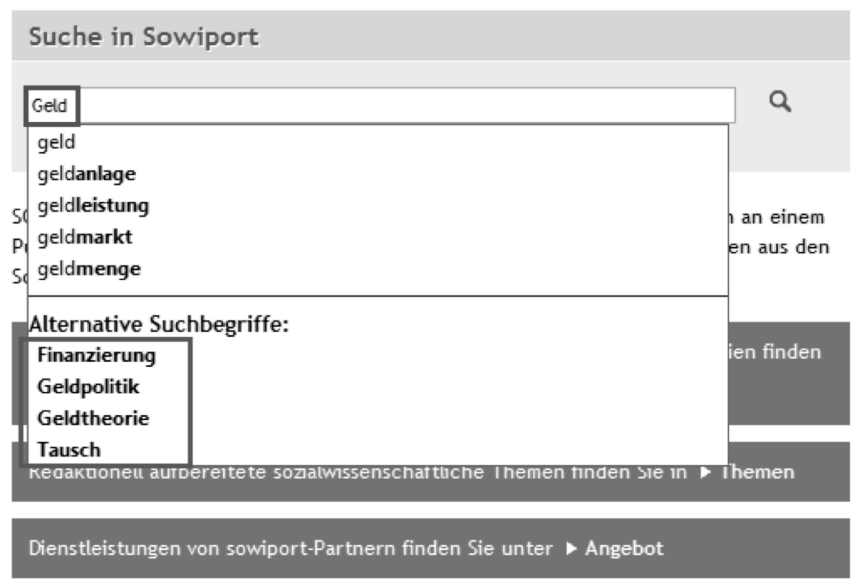

Figure 1: Screenshot of Sowiport. Original query (top frame) and recommended terms (bottom frame).

---

[2] http://www.gesis.org/sowiport

## Automatic Query Expansion

In contrast to a manual user-selected query expansion providers of DL systems could also use our recommendation web service to implement automatic expansion mechanisms. Given a specific query the search engine would automatically expand the query through a certain number of recommended terms (e.g. the five most relevant). This is especially interesting for the recommendation in a broad field where an end-user might not know which expansion to choose. An automatic query expansion would alleviate the user of this problem by combining expansions of the highest relevance.

## Tag Cloud Generator

Another popular use case is a tag or term cloud generator. As described above the STR service calculates the semantic distance between freely chosen input words or phrases and controlled vocabulary terms. A tag cloud may then be used to visualize these distances and provide an intuitive representation for the end-user. See figure 2 for an example of such a term cloud. Terms from the controlled vocabulary that are more relevant to the given query are displayed in a bigger font, thus standing out from the rest of the cloud. With this representation users can refine their search after they have seen (and might be unsatisfied with) the original result list. They would also get information on relative relevance of recommendations instead of getting them in order of relevance alone.

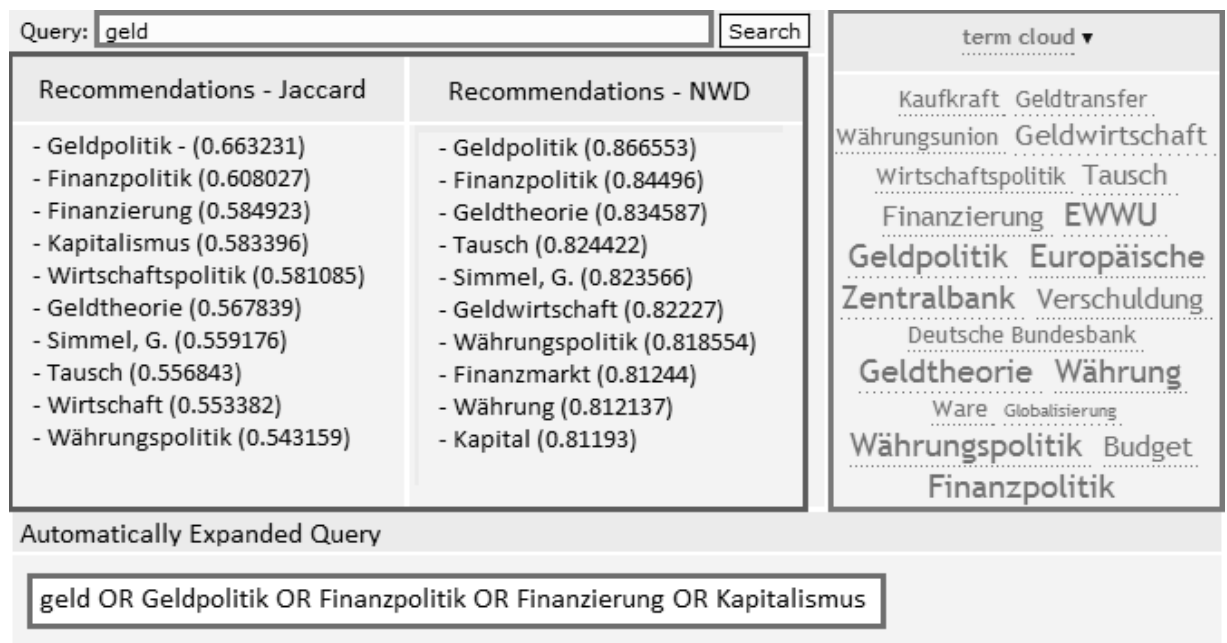

Figure 2: Screenshot from an early prototype presenting Jaccard and NWD (left frame) recommendations, a Mindserver-based term recommendation visualized as a term cloud (right frame) and a proposal for an automatic query expansion (bottom frame) for a given entry term „geld" (top frame).

## Bibliometric Analysis

An additional use case besides the proposed query-supporting mechanisms is to do basic bibliometric analyses on the available data sets. The co-word analyses and classification algorithms that are used to compute the STR can be utilized to gain deeper insights into the underlying semantic structure of the data. What are the most frequent

keywords? What controlled vocabulary terms co-occur with what typical terms in the title? What are emerging terms over time?

Besides rather simple visualizations like tag- or term-clouds and textual output in form of lists or word clusters that only depend on the count of term frequencies, more sophisticated "maps of science" can be created. Examples of such maps are presented on the website of the same title Maps-of-Science [3] or in comparable works of e.g. Leydesdorff & Rafols (2009).

# Technical Workflow and Implementation Details

The description of the technical workflow is based on our own prototype. Up to now [4] this prototype is not open to the public and exists in separate parts that will have to be combined in the near future.

The following section will describe the functionality of the web frontend, what steps are done in pre-processing and the final web service accessible by the user.

## Web Frontend

Our System is accessible through a web frontend that can be viewed as an administration surface. It handles user registration, status mails and also provides rudimentary management and accessibility methods for the domain specific STRs. To make use of their custom STR web service providers of a DL must create an account on our website. The account is secured through the use of a username/password combination. Within this initial registration the contact data of the provider and the OAI-PMH interface of the repository are specified.

After the domain specific STRs are calculated (as described in the following section pre-processing) the results can be tested through the web frontend. For every account there will be different sections that provide useful tools to analyse, access and make use of our STRs. In a first section the different query-supporting mechanisms are previewed. Thus the provider of a DL can get in touch with the way our services work and decide which service fits its needs best. When a decision is made the provider can make use of manuals that that will help to easily embed the desired STR into a repository. In a second section methods for bibliometric analyses are provided. As our project evolves the amount of tools and sections will increase providing a wide variety of tools, methods and support. Figure 2 shows an example of the query-supporting mechanism section.

## Pre-processing and Computation of the Term Suggestions

For the first step of our service, gathering the metadata of a DL, we need a way to provide homogeneity of the underlying data. Also we want to request a minimum of detailed knowledge and interaction from providers of DLs in order to use our service. The OAI-PMH interface that we use to harvest a repositories data satisfies these requirements. It is a commonly used way to provide a repositories metadata. It is a build-in capability of almost any repository software available. Part of the OAI-PMH is the Dublin Core (dc)[5] metadata schema that defines a set of basic metadata like title, authors, subjects and abstract. Figure 3 shows an example of the metadata that is extracted using an OAI-PMH interface.

---

[3] http://mapofscience.com

[4] This article was written in September 2011 – The date of printing is not supposed to be earlier than spring 2012.

```
header:
  identifier : oai:gesis.izsoz.de:19389
  datestamp : 2011-01-10T13:46:00Z
  setSpec : SSOAR

metadata:
  dc:
    identifier: http://nbn-resolving.de/urn:nbn:de:0168-ssoar-193894
    title: How can international donors promote transboundary water management?
    creator: Mostert, Erik
    creator: Deutsches Institut für Entwicklungspolitik gGmbH
    subject: Political science (320)
    subject: Life sciences, biology (570)
    subject: International Relations, International Politics, Development Policy (10505)
    subject: Ecology, Environment (20900)
    subject: Management; Afrika; Entwicklung; Entwicklungsland; Akteur; Wasser
    source: Bonn
    source: DIE Discussion Paper (1860-0441) 8/2005
    description: "This paper discusses how international donors can promote the
        development of transboundary water management. It assumes, first, that
        cooperation will take place whenever the major stakeholders consider cooperation
        to be a better option than non-cooperation. The perceptions and motivations of
        the stakeholders are therefore crucial. Secondly, this paper assumes that the
        major stakeholders are not 'states', but specific groups and individuals:
        individual politicians, sectoral government bureaucracies, regional and local
        governments, farmers, electricity companies, etc. Some of these may be involved
        in the international negotiations themselves, others may be needed to get
        international agreements ratified or implemented, and still others may be
        affected by transboundary water management but lack the means to exert any
        influence." (author's abstract)
    language: English
    rights: Deposit Licence - No Redistribution, No Modifications
    contributor: SSOAR - Social Science Open Access Repository
    date: 10.01.2011 13:46
```

**Figure 3: Sample metadata output from an OAI-PMH interface. This example was taken from the SSOAR repository and formatted to allow better readability.**

Generally in a DL a controlled vocabulary is used to assign a structure to the documents. The vocabulary from which recommendations are chosen is created automatically on basis of the harvested data. We assume that title and abstract (dc:title and dc:description) contain natural language information to describe the content of the document, while (controlled) keywords are encoded in fields like dc:subject. In case of keyword recommendation based on an explicitly controlled vocabulary, the vocabulary can be uploaded and the terms in dc:subject are filtered according to the uploaded list. This is useful if a DL provider wants to exclude certain terms from appearing as recommendation, i.e. if they are too general. Note however that an explicitly specified vocabulary may only be a subset of the actual vocabulary used in creation of the repository. Terms not found in the original vocabulary (and therefore not used in tagging of documents) will never be used as recommendations.

At the moment we provide three different modules basing on different approaches of calculating the recommendations. One is based on machine learning using pLSA and Support-Vector-Machines and implemented by the commercially available software Mindserver. As the underlying software is proprietary we cannot provide a deeper insight of the functionality of this module. The other two modules are based on semantic distance metrics, namely the Normalized Web Distance (NWD) (Cilibrasi & Vitányi, 2009) and the Jaccard Index (Jaccard, 1901). We use terms from titles and abstracts as possible search terms to calculate pairwise distances between these terms and the terms of the controlled vocabulary for which the STR is supposed to be created. While all modules provide recommendations, results may vary depending on the type of module used. Which module provides the best results for a given domain is to be evalua8ted. With increasing experience it might be possible to give recommendations on optimal combinations of STR module and domain type.

---

[5] http://dublincore.org/metadata-basics/

After the metadata has been harvested from the DL, it will be parsed into an internal representation. Stop words will be eliminated and the remaining terms will be stemmed. In the case of the metric based modules all data is inserted into a SQL database to calculate the distances. The Computation of search-term-recommendations is getting more expensive with a growing size of the repository. The Pre-Processing steps will therefore take some time to complete (in the range of hours). When the pre-processing has been completed an email will be sent out. In order to allow all final services to be used in live environments, once the pre-processing phase has been completed, the individual requests for recommendation can be handled in real-time.

The results of the pre-processing will be stored in a database. This database will build the basis of our service. This service can be accessed using a RESTful API (described in the section RESTful API). Figure 4 shows the workflow of our system.

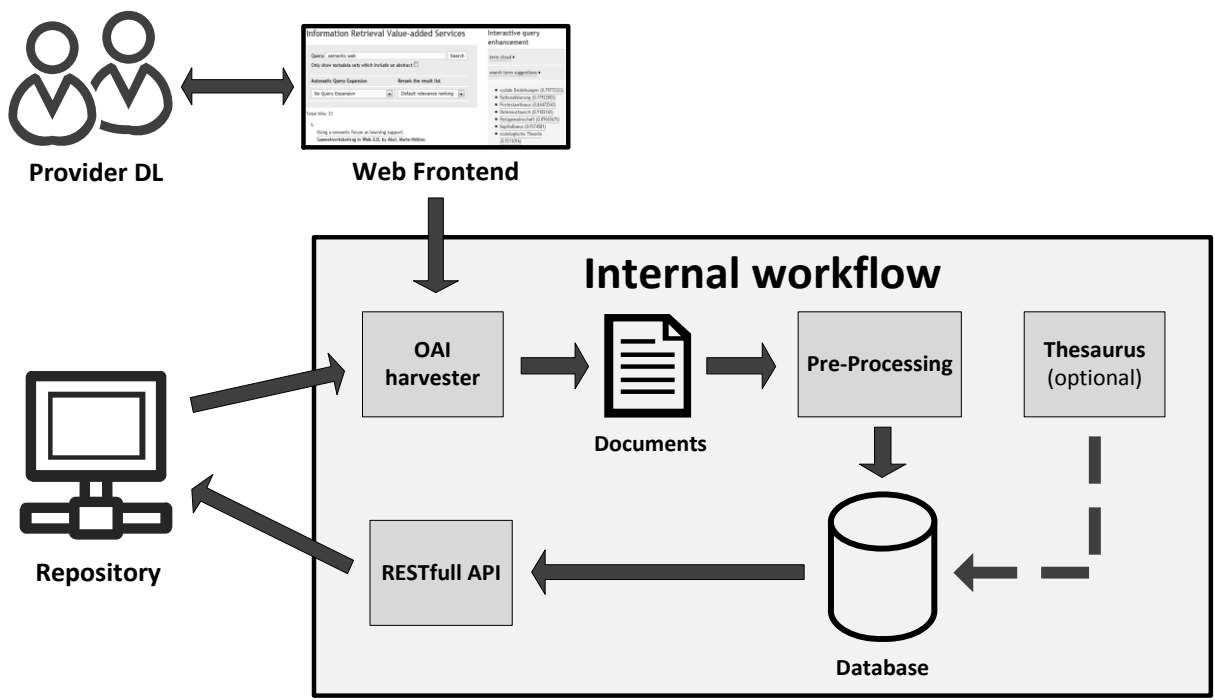

**Figure 4: Idealized workflow for the implementation of a custom STR. A DL provider can register a repository's OAI-PMH interface via a web frontend. The OAI content is harvested and pre-processed and finally written in a database to allow fast access. All computed associations can be accessed through the RESTful API.**

### RESTful API

After specifying information about their data users will be assigned an individual API key. When pre-processing has been completed and a recommendation module was chosen the web services may be accessed through the use of this API key. Use of such a key system is a trade-off between security and comfort of usage. While every person in possession of the key is able to access our services only those in possession of the proper username/password combination may change a profile.

Figure 5 shows an example of the final web service. Users can send a simple HTTP GET request and specify the term or phrase they want recommendations for. In the example the top 10 (based on semantic closeness) results are returned (compare "limit=10" in the address bar) for the search term "Geld" (compare "term=Geld" in address bar).

The result is an XML-style document, containing a list of recommendations, with their name, the confidence or semantic closeness to the search term and some additional information about the vocabulary they are from.

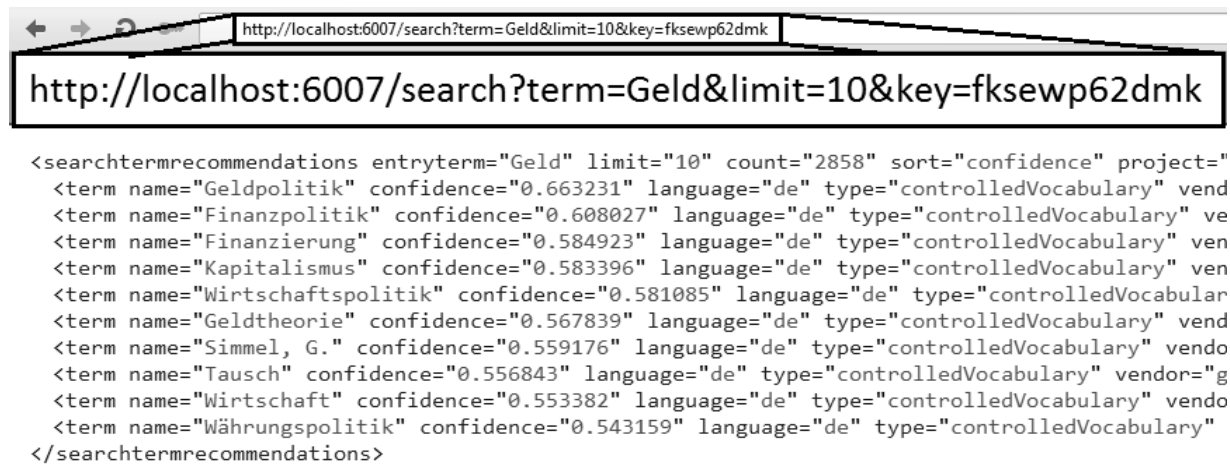

Figure 5: Screenshot from the prototype RESTful web service which returns highly related controlled terms for a given term "geld".

## Discussion and Future Work

In this paper we presented a general system architecture and workflow to build custom web services for so-called Search-Term-Recommenders. These recommender web services can be used to offer an interactive term suggestion mechanism that automatically maps free user input terms onto a domain-specific controlled vocabulary or to do an automatic query expansion. Other implementations like term clouds or a bibliometric analysis tool are possible, too.

In the past comparable systems were not publically available or not customizable for specific DL systems like open access repositories. With the help of the presented system users and operators of such repositories would benefit in terms of usability and interoperability.

The presented approaches and systems are currently being implemented at GESIS in the DFG-funded IRM2[6] project. We gave a first look on the working prototype. Unfortunately this prototype is not publicly available to the print of this article but the general feasibility of the approach has been shown. Within the first public release we would like to see the following features available:

- alternative ways of transmitting the repositories' data (besides OAI-harvesting),

---

[6] http://www.gesis.org/forschung/drittmittelprojekte/projektuebersicht-drittmittel/irm2/

- a more user-friendly way of specifying the user-specific vocabularies, and
- different co-word and categorization algorithms to choose from.

Besides the mentioned points some open topics remain as future work: Additionally to the presented web services we plan to release an open-source licenced Java-based library for external repository developers. We hope that this allows new and inspiring independent solutions. Import and export of the data structure will be available to allow further maintenance of the evaluation results. Within the web interface there will be some visualization included to make the browsing of the results more intuitive.

Please check the projects website for more recent news.

## Acknowledgement

This work was funded by DFG, grant no. SU 647/5-2.